\documentclass[12pt]{article}
\usepackage{chicago}
\if FL 

\newlength{\defitemindent}
     {\begin{list}{}
               {\labelsep=0ex
            \settowidth{\labelwidth}{\hspace{\defitemindent} #1 \hspace{0.5ex}}
            \leftmargin=\labelwidth
            \addtolength{\leftmargin}{\labelsep}
             }}
     {\end{list}}

           {\arraycolsep=0.14em
                      \begin{eqnarray}}
                      {\end{eqnarray}}

\newenvironment{Eqnarray*}%
           {\arraycolsep=0.14em
                      \begin{eqnarray*}}
                      {\end{eqnarray*}}

\fi 

\newtheorem{THEOREM}{Theorem}[section]
\newenvironment{theorem}{\begin{THEOREM}\ }%
                        {\end{THEOREM}}
\newtheorem{LEMMA}[THEOREM]{Lemma}
\newenvironment{lemma}{\begin{LEMMA}\ }%
                      {\end{LEMMA}}
\newtheorem{COROLLARY}[THEOREM]{Corollary}
\newenvironment{corollary}{\begin{COROLLARY}\ }%
                          {\end{COROLLARY}}
\newtheorem{PROPOSITION}[THEOREM]{Proposition}
\newenvironment{proposition}{\begin{PROPOSITION}\ }%
                            {\end{PROPOSITION}}
\newtheorem{DEFINITION}[THEOREM]{Definition}
\newenvironment{definition}{\begin{DEFINITION}\ \rm}%
                            {\end{DEFINITION}}
\newtheorem{CLAIM}[THEOREM]{Claim}
\newenvironment{claim}{\begin{CLAIM}\ \rm}%
                            {\end{CLAIM}}
\newtheorem{EXAMPLE}[THEOREM]{Example}
\newenvironment{example}{\begin{EXAMPLE}\ \rm}%
                            {\end{EXAMPLE}}
\newtheorem{REMARK}[THEOREM]{Remark}
\newenvironment{remark}{\begin{REMARK}\ \rm}%
                            {\end{REMARK}}

\newcommand{\thm}{\begin{theorem}}
\newcommand{\lem}{\begin{lemma}}
\newcommand{\pro}{\begin{proposition}}
\newcommand{\dfn}{\begin{definition}}
\newcommand{\rem}{\begin{remark}}
\newcommand{\xam}{\begin{example}}
\newcommand{\cor}{\begin{corollary}}
\newcommand{\prf}{\noindent{\bf Proof:} \hspace{0.677em}}
\newcommand{\ethm}{\end{theorem}}
\newcommand{\elem}{\end{lemma}}
\newcommand{\epro}{\end{proposition}}
\newcommand{\edfn}{\bbox\end{definition}}
\newcommand{\erem}{\bbox\end{remark}}
\newcommand{\exam}{\bbox\end{example}}
\newcommand{\ecor}{\end{corollary}}
\newcommand{\eprf}{\bbox\vspace{0.1in}}
\newcommand{\beqn}{\begin{equation}}
\newcommand{\eeqn}{\end{equation}}

\newcommand{\clm}{\begin{claim}}
\newcommand{\eclm}{\end{claim}}

\newcommand{\bbox}{\vrule height7pt width4pt depth1pt}

\newcommand{\IR}{\mbox{$I\!\!R$}}

\renewcommand{\phi}{\varphi}

\newcommand{\F}{{\cal F}}

\newcommand{\I}{{\cal I}}

\newcommand{\ol}{\setlength{\itemsep}{0pt}\begin{enumerate}}
\newcommand{\eol}{\end{enumerate}\setlength{\itemsep}{-\parsep}}
\newcommand{\ul}{\setlength{\itemsep}{0pt}\begin{itemize}}
\newcommand{\dl}{\setlength{\itemsep}{0pt}\begin{description}}
\newcommand{\edl}{\end{description}\setlength{\itemsep}{-\parsep}}
\newcommand{\eul}{\end{itemize}\setlength{\itemsep}{-\parsep}}

\newcommand{\commentout}[1]{}

\newcommand{\bi}{\begin{itemize}}
\newcommand{\ei}{\end{itemize}}
\newcommand{\be}{\begin{enumerate}}
\newcommand{\ee}{\end{enumerate}}

\def\rarrowr{\buildrel{\smash{\raise 0.5ex \hbox{$\scriptstyle r$}}} \over
           {\smash{\mathop{\hbox to 1.3em {\rightarrowfill}}}} }

\newcommand{\gammafair}%
{\gamma^{{\it bt}}_{\mbox{\scriptsize{{\it fair}}}}}
\newcommand{\gammafairk}%
{\gamma^{{\it bt}}_{\mbox{\scriptsize{{\it fair,k}}}}}

\newcommand{\eps}{\varepsilon}

\newcommand{\Ifmp}{\mbox{$\I^{\kern 0.1ex \it fm'}$}}

\newtheorem{EXERCISE}{}
{\end{EXERCISE}}
\newtheorem{HARDEX}[EXERCISE]{\llap{*}}
{\end{HARDEX}}
\newtheorem{SUPERHARD}[EXERCISE]{\llap{**}}
{\end{SUPERHARD}}
\newcommand{\oldindex}[1]{}

\usepackage[normalem]{ulem}
\usepackage{color}
\newcommand{\red}{\color{black}}
\newcommand{\black}{\color{black}}
\renewcommand{\sout}{\commentout}

\newcommand{\br}{\mathbf{r}}

\renewcommand{\S}{{\cal S}}

\newcommand{\EU}{{\rm EU}}
\newcommand{\vecS}{{\vec{S}}}   
\newcommand{\vecsig}{\vec{\sigma}}

\newcommand{\stand}[1]{\mbox{\em st}\left (#1 \right )}

\newcommand{\below}{\succ}
\newcommand{\setI}{{\cal I}}

\newcommand{\strat}{{\bf s}}
\newcommand{\hist}{{\bf Z}}

\newcommand{\height}{\mathit{height}}

\begin{document}
\begin{titlepage}
\title{Characterizing Solution Concepts 
in Terms of Common Knowledge of Rationality%
}   
   
   
\author{Joseph Y. Halpern%
\thanks{Supported in part by NSF under grants   
CTC-0208535, ITR-0325453, and IIS-0534064, by ONR under grant
N00014-02-1-0455, by the DoD Multidisciplinary University Research   
Initiative (MURI) program administered by the ONR under   
grants N00014-01-1-0795 and N00014-04-1-0725, and by AFOSR under grants
F49620-02-1-0101 and FA9550-05-1-0055.}   
\\ Computer Science Department \\ Cornell University,   
U.S.A. \\  e-mail: halpern@cs.cornell.edu    
\and    
Yoram Moses%
\thanks{The Israel Pollak academic chair at the Technion; work supported
in part by Israel Science Foundation under grant 1520/11.}\\    
Department of Electrical Engineering\\   
Technion---Israel Institute of Technology\\   
32000 Haifa, Israel\\   
email: moses@ee.technion.ac.il
}

\maketitle
\thispagestyle{empty}
\begin{abstract}   
Characterizations of Nash equilibrium, correlated equilibrium, and
rationalizability in terms of common knowledge of rationality are well
known \cite{Aumann87,BD87a}.
Analogous 
characterizations of sequential equilibrium,
(trembling hand)
perfect equilibrium, and quasi-perfect equilibrium
in $n$-player games are obtained here, 
using results of Halpern
\citeyear{Hal36,Hal36erratum}.
\end{abstract}   
\end{titlepage}

\section{Introduction}
Arguably, the major goal of epistemic game theory is to characterize
solution concepts epistemically.  
%
Characterizations of 
the solution concepts that 
are most commonly 
used in strategic-form
games, namely,
Nash equilibrium, correlated equilibrium, and
rationalizability, in terms of common knowledge of rationality are well
known \cite{Aumann87,BD87a}.  
We show how to get analogous
characterizations
of sequential equilibrium \cite{KW82},
(trembling hand)
perfect equilibrium \cite{Selten75}, and quasi-perfect equilibrium
\cite{vD84} 
for arbitrary $n$-player games,
using results of Halpern \citeyear{Hal36,Hal36erratum}. 

To put our results in context, we start by reviewing the
characterizations of Nash equilibrium, correlated equilibrium, and
rationalizability in Section~\ref{sec:review}.  In
Section~\ref{sec:newresults}, we 
recall 
Halpern's characterizations of
sequential equilibrium and perfect equilibrium, since these play a key
role in our new results.  Halpern's results involve the use of
\emph{nonstandard probability measures}, 
which 
take values in \emph{non-Archimedean fields}.  
We briefly review these as well, and then state and prove the new
characterizations of sequential equilibrium, quasi-perfect
equilibrium, and perfect 
equilibrium in terms of common knowledge of rationality.   
For our results, we need to consider two types of
rationality: \emph{local rationality}, which considers only 
whether
each player's action is a best response at each information set (with
everything else fixed), and \emph{rationality}, which
considers whether his whole strategy from that point on is a best response.
This distinction seems critical when comparing perfect and
quasi-perfect equilibrium (as already noted by van Damme
\citeyear{vD84}); interestingly, it is not critical when it comes to
sequential equilibrium. 
We  compare our results to those of Asheim and 
Perea
\citeyear{AP05}, who 
provide a characterization of sequential equilibrium
and quasi-perfect equilibrium for 2-player games in
terms of common knowledge of rationality similar in spirit to ours. 
We conclude
in Section~\ref{sec:discussion} with a discussion of the use of common
knowledge of rationality in characterizing solution concepts.

\section{A review of earlier results}\label{sec:review}
To explain our results, we briefly review the 
earlier results
on characterizing solution concepts in strategic-form games terms of
common knowledge (see \cite{DS15} for a more comprehensive survey).  
We assume that the reader is familiar with standard solution concepts
such as Nash equilibrium, correlated equilibrium, and rationalizability;
see \cite{OR94} for a discussion.  
Let $\Gamma = (N,\S,(u_i)_{i \in N})$ be a 
finite strategic-form game, where $N =
\{1,\ldots, n\}$ is the set of players, $\S= \times_{i \in N}\S_i$ is
a finite set of strategy profiles, and $u_i : \S \rightarrow \IR$ is player
$i$'s utility function.  
For ease of exposition, we assume that $\S_i\cap \S_j=\emptyset$ for
$i\ne j$. 

Let a \emph{model of $\Gamma$} be a tuple $M =
(\Omega,\strat,(\Pr_i)_{i \in N})$, 
where $\Omega$ is a set of 
states of~$\Gamma$, 
$\strat$~associates with each state
$\omega \in \Omega$ a pure strategy profile $\strat(\omega) \in \S$, and
$\Pr_i$ is a probability distribution on $\Omega$, describing $i$'s initial
beliefs.%
\footnote{For simplicity, we assume in this paper that $\Omega$ is
finite, and all subsets of $\Omega$ are measurable.}
Let $\strat_i(\omega)$ denote player $i$'s strategy in the
profile $\strat(\omega)$, and let $\strat_{-i}(\omega)$ denote the strategy
profile consisting of the strategies of 
all
players other than $i$.  

For $S\in\S_i$, let 
$[S] = \{\omega \in \Omega: \strat_i(\omega) = S\}$
be the set of states at which player~$i$ chooses strategy~$S$. 
Similarly, let 
$[\vecS_{-i}] = \{\omega \in \Omega: \strat_{-i}(\omega) =
\vecS_{-i}\}$
and $[\vecS] = \{\omega \in \Omega: \strat(\omega) = \vecS\}$.
For simplicity, we assume that $[\vecS]$ is measurable for all strategy
profiles $\vecS$, and that 
$\Pr_i([S_i]) > 0$ 
for all strategies $S_i \in
\S_i$ and all players $i \in N$.

As usual, we say that a player is \emph{rational} at state~$\omega$ 
(in a model~$M$ of~$\Gamma$) 
if his strategy 
at~$\omega$ 
is a best response 
in~$\Gamma$ given his beliefs at $\omega$.
We view
$\Pr_i$ as $i$'s prior belief, intuitively, before $i$ has been assigned
or has chosen a strategy.  We assume that $i$ knows his strategy at
$\omega$, and that this is all that $i$ learns in going from his prior
knowledge to his knowledge at $\omega$, 
so his beliefs at $\omega$ are the result of conditioning 
$\Pr_i$ on $[\strat_i(\omega)]$.%
\footnote{While this arguably is a reasonable assumption for
  strategic-form games, when we move to extensive-form games, agents
  will be able to learn more in the course of a game.}
Given our assumption that 
$\Pr_i([\strat_i(\omega)]) > 0$, the conditional probability
~$\Pr_i \mid [\strat_i(\omega)]$ ~is well defined.

Note that we can view $\Pr_i$ as 
inducing
a probability 
$\Pr_i^{\S}$
on strategy profiles 
$\vecS\in\S$
by simply 
taking $\Pr_i^{\S}(\vecS) =  \Pr_i([\vecS])$; we similarly define
$\Pr_i^{\S}(S_i) =  \Pr_i([S_i])$ and 
$\Pr_i^{\S}(\vecS_{-i}) =  \Pr_i([S_{-i}])$.
Let $\Pr_{i,\omega}^{\S} = \Pr_i^{\S} \mid \strat_i(\omega)$.
Intuitively, at state $\omega$, player $i$ knows his strategy
$\strat_i(\omega)$, so his distribution $\Pr_{i,\omega}^{\S}$ on
strategies at $\omega$  is the result of conditioning
his prior distribution  on strategies~$\Pr_i^{\S}$ on this information.

Formally, $i$ is \emph{rational at $\omega$} if,
for all strategies $S \in \S_i$, we have that 
$$\sum_{\vecS_{-i}' \in \S_{-i}}
{\Pr}_{i,\omega}^{\S}(\vecS_{-i}')u_i(\strat_i(\omega),\vecS_{-i}') \ge  
\sum_{\vecS_{-i}' \in \S_{-i}}
{\Pr}_{i,\omega}^{\S}(S_{-i}')u_i(S,\vecS'_{-i}).$$ 
We say that player $i$ is \emph{rational in model $M$} if $i$ is rational at
every state $\omega$ in $M$.  
Finally, we say that \emph{rationality 
is
common knowledge in $M$} if all players are rational 
at
every state of
$M$.%
(Technically, our definition of rationality being common knowledge in
$M$ means that rationality is \emph{universal} in $M$
(i.e., true at {\em all} states in $M$), and 
thus, in particular, 
common knowledge at all
states in $M$ according to the standard definition of common knowledge
at a state (cf., \shortcite{FHMV}).  
While common knowledge of rationality at a state does not imply that rationality
is universal in general,  
in the models that we 
focus on in this paper, the two notions coincide.)

With this background, we can state Aumann's \citeyear{Aumann87}
characterization of Nash equilibrium.  As usual, we can identify a mixed
strategy 
profile $\vecsig$ in $\Gamma$ with a distribution~$\Pr_{\vecsig}$
on~$\S$; the distribution  
$\Pr_{\vecsig}$ can be viewed as a crossproduct $\times_{i \in N}
\Pr_{\sigma_i}$ (where $\Pr_{\sigma_i}$ is a distribution on $\S_i$).%
\footnote{We consistently use $S$, possibly subscripted, to denote a
pure strategy, while~$\sigma$, possibly 
subscripted or with a prime,
denotes a mixed strategy.}
Let $\Sigma_i$ denote the set of mixed strategies for player $i$.

\thm\label{thm:charNE} $\vecsig$ is a Nash equilibrium of
$\Gamma$ iff there exists a model $M = (\Omega,\strat, (\Pr_i)_{i \in
N})$ of $\Gamma$  where rationality is common knowledge such that $\Pr_i
= \Pr_j$ for 
all $i, j \in N$ and $\Pr_i^{\S} = \Pr_{\vecsig}$ for all $i \in N$. 
\ethm

The fact that $\Pr_i = \Pr_j$ for all $i, j \in N$ means that there is a
common prior.  Because~$\Pr_{\vecsig}$ has the form of a cross-product,
the fact 
that $\Pr_i^{\S} = \Pr_{\vecsig}$ means that $i$'s beliefs about other players'
strategies is independent of the state; 
that is, $\Pr_i^{\S} \mid \strat_i(\omega)$ marginalized to $\S_{-i}$ is 
independent of $\omega$.%
\footnote{Aumann and
Brandenburger \citeyear{AB95} show that common knowledge of rationality
is not required for~$\sigma$ to be a Nash equilibrium.  This is not a
contradiction to Theorem~\ref{thm:charNE}, which simply says that~$\sigma$ is a Nash equilibrium iff there
exists a model $M$ describing the beliefs of the players where rationality
is common knowledge.  There may be other models where the players play
$\sigma$ and rationality is not common knowledge.}

Theorem~\ref{thm:charNE} is actually a special case of Aumann's
\citeyear{Aumann87} characterization of correlated equilibrium.  
Recall that we can think of a correlated equilibrium of $\Gamma$ as a
distribution~$\eta$ on $\S$.  Intuitively, $\eta$ is a correlated
equilibrium if, when a mediator chooses a strategy profile $\vecS$
according to $\eta$ and tells each player $i$ his component $S_i$ of
$\vecS$, then 
playing $S_i$ is a best response for~$i$.
This intuition is formalized in Aumann's theorem:

\thm\label{thm:charCE} $\eta$ is a correlated equilibrium of $\Gamma$
iff there exists a model $M = (\Omega,\strat, (\Pr_i)_{i \in
N})$ of $\Gamma$  where rationality is common knowledge such that $\Pr_i^\S
= \eta$ for 
all $i \in N$.
\ethm

Theorems~\ref{thm:charNE} and~\ref{thm:charCE} show 
that the difference between
correlated equilibrium and Nash equilibrium can be understood as saying
that, with correlated equilibrium, the common prior does not have to
be
a cross-product, so that 
a player $i$'s beliefs may vary, for different choices of strategy.
Of course, if 
the prior {\em is}
a cross-product, then the correlated equilibrium is
also a Nash equilibrium.  With correlated equilibrium, as with Nash
equilibrium, there is a common prior.

We complete the review of 
characterizations of 
solution concepts in
strategic-form
games in terms of common knowledge of rationality with
the following characterization of correlated
rationalizability (where a player can believe that other players' strategies
are correlated), due to Brandenburger and Dekel \citeyear{BD87a}:

\thm\label{thm:rationalizable} 
$S_j$ is a (correlated) rationalizable strategy for player~$j$ in a
game~$\Gamma$ iff    
there exists a model $M = (\Omega,\strat, (\Pr_i)_{i \in N})$ of
$\Gamma$ where
rationality is common knowledge and a state 
$\omega\in\Omega$ such that \mbox{$\strat_j(\omega) = S_j$}.
\ethm

Note that the characterization of rationalizability does not require 
the players to have a common prior.  

\section{Characterizing sequential equilibrium and perfect
  equilibrium}\label{sec:newresults} 
Our goal is to characterize sequential equilibrium and
perfect equilibrium in 
finite
extensive-form games with perfect recall in
terms of common knowledge of rationality.   
We assume that the reader is
familiar with the standard definitions of extensive-form games of perfect
(trembling hand) perfect
equilibrium, quasi-perfect equilibrium, and sequential
 equilibrium. 
Our characterizations make essential use of non-epistemic 
 characterizations of
sequential and perfect equilibrium using nonstandard probability
\cite{Hal36,Hal36erratum}.   
We briefly review these results here.

One of the issues that 
the definitions of sequential and perfect equilibrium need to
deal with are probability zero events, specifically, those corresponding to 
information sets that are off the equilibrium path. Halpern
\citeyear{Hal36,Hal36erratum} presents a novel way to 
approach this issue in the context of games, by making use of
nonstandard probability measures, which we now describe. 

{\em Non-Archimedean fields} are fields
that include the real  numbers $\IR$ as a subfield, and also contain 
{\em infinitesimals\/}, which are positive numbers that are strictly
smaller than any positive real number.   
The smallest such non-Archimedean field,
commonly denoted~$\IR(\eps)$, 
is the minimal field generated by
adding to the reals a single infinitesimal, denoted by~$\eps$.%
\footnote{The construction of $\IR(\eps)$ apparently goes back to
Robinson \citeyear{Robinson73}.} 
$\IR(\eps)$ consists of all the rational expressions 
$f(\eps)/g(\eps)$, where $f(x)$ and $g(x)$ are polynomials with
real coefficients and $g(0) \ne 0$.  It is easy to see 
that 
this gives us a
field that includes the reals and $\eps$.  We can place an order $<$
on the elements of $\IR(\eps)$ by taking $0 < \eps < 1/r$ for all reals
$r > 0$, and extending to all of $\IR(\eps)$ by assuming
that standard properties of the reals 
(e.g., that
$r^2 < r$ if $0 < r < 1$)
continue to hold.  
 Thus, $0 < \cdots < \eps^3 < \eps^2 < \eps$ holds, for all real
 numbers $r > 0$ we have that $1/\eps > r$,  
and so on.  (We can use formal division to
identify $f(\eps)/g(\eps)$ with a power series of the form $a_0 + a_1
\eps + a_2 \eps^2 + \cdots$; this suffices to guide how the order $<$
should be extended to quotients $f(\eps)/g(\eps)$.)

The field $\IR(\eps)$ does not suffice
for our purposes.
In this paper we will be interested in non-Archimedean fields
$\IR^*$ that are \emph{elementary extensions} of the standard reals.
This means that $\IR^*$ is an ordered field that
includes the real numbers, at least one 
infinitesimal~$\eps$,
and is \emph{elementarily equivalent to the
field of 
real numbers}.  The fact that $\IR^*$ and $\IR$ are elementarily equivalent
means that every formula 
$\varphi$
that can be expressed in first-order logic and
uses the function symbols $+$ and $\times$ (interpreted as addition and 
multiplication, respectively), and 
constant symbols~$\br$ standing for particular real numbers 
(the underlying language contains a constant symbol~$\br$ for each
real number $r\in\IR$) is true in $F$ iff  $\varphi$ is true in $\IR$.
%
We call such a field a \emph{normal} non-Archimedean field. 
Thus, for example, every odd-degree polynomial has a
root in 
a 
normal non-Archimedean field~$\IR^*$ 
since this fact is true in $\IR$ and can be expressed in first-order logic.
Note that 
$\IR(\eps)$ 
is not a normal non-Archimedean field.  For
example, one property of the reals expressible in first-order logic is
that every positive number has a
square root.  However, $\eps$ does not have a square root in
$\IR(\eps)$.
For the results of this paper, we do not have to explicitly describe
a normal non-Archimedean field; it suffices that one exists.
The existence of normal non-Archimedean fields is well known, and
follows from the fact that first-order logic is \emph{compact};
see \cite{Enderton}.%
\footnote{There is a natural extension of
$\IR(\eps)$ called $\IR^*(\eps)$ that is normal.  As shown by
\cite{Hal36,Hal36erratum}, 
Theorems~\ref{thm:thpe} and~\ref{thm:seqeq} could be strengthened 
to use $\IR^*(\eps)$ rather than an existentially quantified normal
non-Archimedean field.}

Given a normal non-Archimedean field $\IR^*$, we call the elements 
of~$\IR$ the {\em standard} reals in~$\IR^*$, and those of
$\IR^*\setminus\IR$ the {\em nonstandard} reals. A nonstandard
real~$b$ is {\em finite} if $-r < b < r$ for some standard real $r > 0$. 
If $b\in\IR^*$ is a finite nonstandard real, then
$b = a + \eps$, where $a$ is the unique standard real number
closest to $b$ and $\eps$
is an infinitesimal.  Formally, $a = \inf\{r \in \IR: r > b\}$ and
$\eps = b-a$; it is easy to check that $\eps$ is indeed an infinitesimal.  
We call $a$ the \emph{standard part} of $b$, and denote it $\stand{b}$.

A nonstandard probability measure $\Pr$ on $\Omega$ just
assigns each event in $\Omega$ an element in $[0,1]$ in some (fixed)
non-Archimedean field $\IR^*$.
Note that $\Pr(\Omega)=1$, just as with standard
probability measures. We require $\Pr$ to be finitely additive. 
Recall that, for 
the purposes of this paper, we 
restrict attention to finite state spaces $\Omega$.  This allows us to
avoid having to define an 
analogue of countable additivity 
for nonstandard probability measures.
Given a nonstandard probability measure $\nu$, we can
define the standard probability measure~$\stand{\nu}$ by taking
$\stand{\nu}(w) = \stand{\nu(w)}$.     
Two possibly nonstandard distributions $\nu$ and~$\nu'$
{\em differ   infinitesimally} if $\stand{\nu}=\stand{\nu'}$
 (i.e., for all events $E$, 
 the probabilities 
 $\nu(E)$ and $\nu'(E)$ differ by at most
an infinitesimal, so $\stand{\nu(E) - \nu'(E)} = 0$).
If a nonstandard distribution assigns a positive (possibly
infinitesimal) probability to every possible outcome in a game, then
there is no technical problem in conditioning on such outcomes. 
Moreover, every standard probability measure differs infinitesimally from a
nonstandard probability measure that assigns positive probabilities to
all outcomes.

\commentout{
We take a \emph{standard} mixed strategy profile $\vecsig$ to be one
where each 
strategy
$\sigma_i$ defines a standard probability over pure
strategies for~$i$;
in a \emph{nonstandard} mixed strategy profile~$\vecsig$, the
strategies $\sigma_i$ can define a nonstandard probability over pure 
strategies. 
Two mixed strategy profiles {\em differ infinitesimally} if the
distributions on strategies that they induce differ infinitesimally. 
}
A {\em behavioral} strategy $\sigma$ for player $i$ in an extensive-form game
associates with each information set $I$ 
for player $i$
a distribution $\sigma(I)$ over
the actions that can be played at $I$.
We allow $\sigma(I)$ to be a nonstandard probability distribution.
We say that $\sigma$ is \emph{standard} if $\sigma(I)$ is standard for
all information sets $I$ for player $i$. 
Two behavioral strategy $\sigma$ and $\sigma'$ for player $i$
\emph{differ infinitesimally} if, for all information sets $I$ for
player $i$, the distributions $\sigma(I)$ and $\sigma'(I)$ differ
infinitesimally.  Two strategy profiles $\vecsig$ and $\vecsig'$ differ
infinitesimally if $\sigma_i$ and $\sigma'_i$ differ infinitesimally
for $i = 1,\ldots, n$.  
%
We say that a 
behavioral strategy $\sigma$ is \emph{completely mixed} if it
assigns positive (but possibly infinitesimal) probability to every
action at every information set. 

A behavioral
strategy profile in an extensive-form game induces a probability on
\emph{terminal histories} of the game (i.e., histories that start at the root
of the game tree and 
end
at a leaf).
Let $Z_\Gamma$ be the set of terminal histories in a game~$\Gamma$.  
(We omit 
explicit mention of the game~$\Gamma$ 
if it is clear from context or irrelevant.)
Given a behavioral strategy profile $\vec{\sigma}$
for~$\Gamma$, 
let $\Pr_{\vec{\sigma}}$ be the probability on terminal histories
induced by $\vec{\sigma}$.
Thus, $\Pr_{\vec{\sigma}}$ is a distribution on pure strategy profiles
if $\vec{\sigma}$ is a mixed strategy profile, and a distribution on histories
if $\vec{\sigma}$ is a behavioral strategy profile in an extensive-form
game.  We hope that the context 
will 
disambiguate the notation.
%
Since we can identify a partial history with
the terminal histories that extend it, $\Pr_{\vecsig}(h)$ and
$\Pr_{\vecsig}(I)$ are well defined for a partial history $h$ and an
information set $I$.
Recall that in an extensive-form game $\Gamma$, each player $i$'s
utility function is defined on $Z_\Gamma$. 
A \emph{belief system} \cite{KW82} 
is a function~$\mu$ 
that associates with each information set $I$ a
probability, denoted $\mu_I$, 
on the histories in $I$.  Given a 
behavioral strategy $\vecsig$ and a belief system $\mu$ in an
extensive-form game $\Gamma$, let
$$\EU_i((\vec{\sigma},\mu)\mid I)=\sum_{h\in I}\sum_{z\in
Z}\mu_I(h){\Pr}_{\vec{\sigma}}(z\mid h)u_i(z).$$
Thus, the expected utility of $(\vec{\sigma},\mu)$ conditional on
reaching~$I$ captures the expected payoff to player~$i$  
if~$I$ is reached via the distribution $\vec{\sigma}$ and from that
point on the game is played according to~$\mu$.  
Intuitively, 
this expected utility
captures what~$i$ can expect to receive if~$i$ changes 
its strategy at information set~$I$.

Finally, if $\vecsig$ is a completely-mixed
behavioral strategy profile, let $\mu^{\vecsig}$ be the belief system
determined by $\vecsig$ in the obvious way:
$$\mu^{\vecsig}_I(h) = {{\Pr}_{\vecsig}}(h\mid I).$$

\dfn 
%
Fix a game~$\Gamma$.
Let  $I$ be an
information set for player $i$, let $\vec{\sigma}'$ be a
completely-mixed behavioral strategy profile, and let $\eps \ge 0$. 
Then we say that
$\sigma_i$
is an \emph{$\eps$-best  
response to $\vec{\sigma}'_{-i}$ for $i$ conditional on having reached
$I$ using $\vec{\sigma}'$} if, for every 
strategy~$\tau_i$
for player $i$,
we have 
that
\begin{equation}
\label{ebest-eq}
\EU_i(((\sigma_i,\vec{\sigma}'_{-i}),\mu^{\vecsig'}_I) \mid I) \ge
\EU_i(((\tau_i,\vec{\sigma}'_{-i}),\mu^{\vecsig'}_I) \mid I) - \eps.
\end{equation}
The strategy 
$\sigma_i$ 
is an \emph{$\eps$-best  
response for $i$ relative to $\vec{\sigma}'$}  if
$\sigma_i$ is an $\eps$-best  
response to $\vec{\sigma}'_{-i}$ for $i$ conditional on having reached
$I$ using $\vec{\sigma}'$  for all information sets $I$ for 
$i$.
\edfn
Observe that in Equation~\ref{ebest-eq} the probability of reaching~$I$
on both sides of the inequality depends only on~$\vec{\sigma}'$ (via
$\mu^{\vecsig'}_I$) and not on~$\tau_i$. Thus, $\tau_i$ only influences
player~$i$'s behavior after~$I$ has been reached.

{\red{Given an information set $I$ for player $i$,
let $A_I$ be the set of actions 
available to~$i$ at histories in~$I$.%
\footnote{As is standard, we assume that the same set of actions is
  available to $i$ at all histories in $I$.}
%
As usual, we take
$\Delta(A_I)$ to be the set of probability measures on $A_I$.  Note
that if $\sigma_i$ is a behavioral strategy for player $i$ then, by
definition, $\sigma_i(I) \in \Delta(A_I)$.  } }

\dfn 
\label{def:local-best}
\red{If $\eps \ge 0$ and $I$ is an
information set for player $i$ that is reached with positive probability
by $\vec{\sigma}'$, then  $a \in \Delta(A_I)$ 
is a \emph{local $\eps$-best  
response to $\vec{\sigma}'_{-i}$ for $i$ conditional on having reached
$I$ using $\vec{\sigma}'$} if, for all $a' \in \Delta(A_I)$, we have
that 
\begin{equation}\label{eq2}
\EU_i(((\sigma_i'[I/a],\vec{\sigma}'_{-i}),\mu^{\vecsig'}_I) \mid I) \ge
\EU_i(((\sigma_i'[I/a'],\vec{\sigma}'_{-i}),\mu^{\vecsig'}_I) \mid
I) - \eps,
\end{equation}
where $\sigma_i'[I/a']$ is the
behavioral strategy that agrees with $\sigma_i'$ except possibly at
information set $I$, and $\sigma_i'[I/a'](I) = a'$.}
The strategy $\sigma_i$ 
is 
a 
\emph{local $\eps$-best  
response for $i$ relative to $\vec{\sigma}'$}  if
$\sigma_i(I)$ is a local $\eps$-best  
response to $\vec{\sigma}'_{-i}$ for $i$ conditional on having reached
$I$ using $\vec{\sigma}'$  for all information sets $I$ for $i$.
%
The strategy $\vec{\sigma}_i$ is a \emph{(local) best  
response for $i$ relative to~$\vec{\sigma}'$} (resp., \emph{(local) best
response for $i$ conditional on having reached
$I$ using $\vec{\sigma}'$}) if $\sigma_i$ is  
a 
(local) 0-best response 
for $i$ relative to $\vec{\sigma}'$ (resp., (local) 0-best
response for $i$ conditional on having reached $I$).
\edfn
Thus, with local best responses, we consider the best \emph{action} at
an information set; with (non-local) best responses, we consider the
best continuation strategy.

\commentout{
\begin{definition}\cite{Hal36erratum}
If $\eps\ge 0$ and $I$ is an information set for $i$, then $\sigma_i$ is
an {\em $\eps$-best 
response for player $i$ relative to $\vec{\sigma'}$ conditional on
having reached $I$} if, for all strategies $\tau_i$ for player $i$, 
$$\EU_i((\sigma_i',I,\sigma_i),\vec{\sigma}_{-i}') \ge 
\EU_i((\vec{\sigma_i',I,\tau_i),\vec{\sigma}_{-i}'}) \ge \eps.$$
Strategy $\sigma_i$ is an \emph{$\eps$-best response for $i$ relative to
$\vec{\sigma}'$} if it an $\eps$-best response for $i$ relative to
$\vec{\sigma}'$ for all information sets $I$ for player  $i$.  
Strategy $\sigma_i$ is a \emph{best response for $i$ relative to
$\vec{\sigma}'$ (at information set $I$)} if it a 0-best response for $i$
relative to $\vec{\sigma}'$ (at information set $I$).
\end{definition}

Note that in the definition of an $\eps$-best response for $i$ relative
to $\vec{\sigma}'$ at information set $I$, player $i$ is assumed to use
$\sigma_i'$ up to information set $I$, and then to switch to $\sigma_i$ at
and below $I$.  That is why we must talk about a best response
\emph{relative to $\vec{\sigma}'$}: we assume that $i$ uses $\sigma'_i$
up to $I$, and the remaining players use $\vec{\sigma}'_{-i}$ throughout.
}

%
%
%
Halpern~\citeyear{Hal36,Hal36erratum} characterizes perfect 
equilibrium using non-Archimedean fields and local best responses as follows: 

\thm
\label{thm:thpe} 
Let $\Gamma$ be a finite extensive-form game with perfect recall. Then 
the (standard) behavioral strategy profile 
$\vecsig=(\sigma_1,\ldots,\sigma_n)$ is a 
perfect equilibrium of $\Gamma$ iff there exists a 
normal non-Archimedean field $\IR^*$ and a 
nonstandard
completely-mixed 
behavioral
strategy profile $\vec{\sigma}'$ 
with probabilities in $\IR^*$
that differs
infinitesimally from $\vec{\sigma}$ 
\red{such that, for each player $i = 1, \ldots, n$ and each information set
$I$ of player $i$, $\sigma_i(I)$ is a local best response for $i$}
relative to $\vecsig'$.
\ethm

Roughly speaking, Theorem~\ref{thm:thpe} shows that we can replace the
sequence of strategies converging to $\vec{\sigma}$ considered in
Selten's definition of perfect equilibrium by a single nonstandard
completely-mixed 
strategy that is infinitesimally close to
$\vec{\sigma}$.  
%
Considering a completely-mixed strategy 
guarantees that all information sets are reached with positive
probability, and thus allows us to define best responses conditional on
reaching 
an information set, for 
every information set.    

We can obtain a characterization of quasi-perfect equilibrium by
requiring that $\sigma_i$ be a best response for $i$ rather than a local best
response \cite{Hal36,Hal36erratum}.%
\footnote{The characterization of perfect equilibrium given in
  \cite{Hal36} involved best responses.  In \cite{Hal36erratum}, it
  was pointed out that this was incorrect; $\sigma_i$ needed to be a
  \emph{local} best response to get a characterization of perfect
  equilibrium, but taking it to be a best response gave a
  characterization of quasi-perfect equilibrium.}
As we said earlier, the fact that the key difference between perfect
equilibrium and 
quasi-perfect equilibrium is that local best responses were required
for the former and best responses were required for the latter 
was already stressed by van Damme~\citeyear{vD84} in his original
definition of 
quasi-perfect equilibrium.
\thm {\rm \cite{Hal36,Hal36erratum}} \label{thm:quasi} 
Let $\Gamma$ be a finite extensive-form game with perfect recall. Then 
the (standard) behavioral strategy profile 
$\vecsig=(\sigma_1,\ldots,\sigma_n)$ is a 
quasi-perfect equilibrium of $\Gamma$ iff there exists a 
normal non-Archimedean field $\IR^*$ and a 
nonstandard completely-mixed behavioral
strategy profile $\vec{\sigma}'$ 
with probabilities in $\IR^*$
that differs
infinitesimally from $\vec{\sigma}$ 
\red{such that, for each 
player $i = 1, \ldots, n$, the strategy~$\sigma_i$ 
is a best response for $i$}
relative to $\vecsig'$.
\ethm

{\red{Finally, we can 
obtain
a characterization of sequential equilibrium by requiring that~$\sigma_i$ be an 
$\eps$-best 
response for $i$ to $\vecsig'$ rather
than a local best response as in Theorem~\ref{thm:thpe}, 
or a 
best response as in Theorem~\ref{thm:quasi}.
It can be
shown if $\eps$ is an infinitesimal, then there exists an
infinitesimal $\eps'$ such that an $\eps$-local best response relative
to $\vecsig'$ is actually an 
$\eps'$-best response (see Lemma~\ref{localtoglobal}), so, as we would expect, the
requirement for sequential equilibrium is actually a weakening
of the requirements for both perfect and quasi-perfect equilibrium.}}


\thm {\rm \cite{Hal36,Hal36erratum}}\label{thm:seqeq}
Let $\Gamma$ be a finite extensive-form game with perfect recall. Then there 
exists a belief system $\mu$ such that the 
assessment $(\vecsig,\mu)$ is a sequential equilibrium of
$\Gamma$ iff there exist a normal non-Archimedean field $\IR^*$,
an infinitesimal $\eps \in\IR^*$, and a nonstandard 
completely-mixed 
behavioral
strategy profile~$\vecsig'$ with probabilities in $\IR^*$
that differs infinitesimally from $\vec{\sigma}$ such  
that~$\sigma_i$ is an $\eps$-best response for $i$ relative to $\vecsig'$, 
for each player $i=1,\ldots,n$.
\ethm
Our epistemic characterizations are based on
Theorems~\ref{thm:thpe}, \ref{thm:quasi}, and~\ref{thm:seqeq}.
Given a finite extensive-form game $\Gamma$, 
we take a model $M$ of $\Gamma$ to be a tuple 
$(\Omega,\hist, (\Pr_i)_{i \in N})$ 
where, as before, $\Omega$ is a 
finite
set of states and $\Pr_i$ is a
(possibly nonstandard) probability distribution on $\Omega$.  Now
$\hist$ is a function that 
associates with each state $\omega \in \Omega$ a terminal history 
in~$\Gamma$, denoted 
$\hist(\omega)$.
The distribution $\Pr_i$ on states induces 
a distribution $\Pr_i^Z$ on terminal histories in the obvious way.
A model $M = (\Omega,\hist,(\Pr_i)_{i \in N})$ of the game~$\Gamma$ is
\emph{compatible with} 
a behavioral strategy profile 
$\vecsig$ if
$\Pr^Z_1 = \cdots = \Pr^Z_n =
\Pr_{\vecsig}$.  

We now define two notions of rationality, corresponding to the types
of best response considered above: local best response and best
response.  
To be consistent with
the type of response considered, we
call these \emph{local rationality} and 
\emph{rationality}.  
Both
notions have been considered in the
literature, although different terms have been used.  Arieli and
Aumann \citeyear{AA15} use the terms 
\emph{action rationality} and \emph{utility maximization} instead of
``local rationality'' and ``rationality''.  

\dfn
Fix $\eps > 0$ and a model $M$ compatible with a
completely-mixed strategy profile $\vecsig'$. 
Player $i$ is \emph{$\eps$-locally rational at state
$\omega$} if, for
each information set $I$ for player~$i$, 
if some history $h \in I$ is a prefix of $\hist(\omega)$, player $i$ 
plays action $a$ after $h$ in
$\hist(\omega)$, 
and 
$\stand{\sigma'_i(I)(a)} > 0$, 
then $a$ is a local $\eps$-best response to $\vec{\sigma}_{-i}$ for
$i$ conditional on having reached $I$ using $\vec{\sigma}'$.
Player $i$ is 
\emph{locally rational} at $\omega$ if he is 0-locally rational at $\omega$.
Player~$i$ is \emph{$\eps$-rational at state
$\omega$} if, for each information set $I$ for player $i$, 
if some history $h \in I$ is a prefix of $\hist(\omega)$, then
$\stand{\sigma_i'}$ is an $\eps$-best response to 
$\vec{\sigma}_{-i}$ for
$i$ conditional on having reached $I$ using $\vec{\sigma}'$.
Player $i$ is \emph{rational at state $\omega$} if he is $0$-rational
at~$\omega$.
\edfn

Note that in the definition of local rationality at $\omega$, we do not 
require that the action played by $i$ 
at a
prefix of $\hist(\omega)$ be a local best
response if that action is played with only infinitesimal probability.
Similarly, in the definition of rationality, we require
$\stand{\sigma_i'}$ to be a best response, not $\sigma_i'$, since we
are ultimately interested in $\stand{\sigma_i'}$.  
Also note that we define rationality only in models that are compatible
with a completely-mixed behavioral strategy profile.  This ensures that
the expected utility conditional on $I$ is well defined for each
information set $I$. 
 We could, of
course, try to define rationality more generally, but the extra work
would not be relevant to the results of this paper.  

\commentout{
It is \emph{almost common knowledge that players are 
$\eps$-rational} 
in a model $M$ if all players are $\eps$-rational
at every state $\omega$ in $M$ such that $\stand{\Pr_i(\omega)}>0$ for
some $i \in N$.
Thus, 
almost common knowledge of $\eps$-rationality holds 
in $M$
if the
probability of all players not being $\eps$-rational is infinitesimal.
}


\sout{
In a finite extensive-form game~$\Gamma$ with perfect recall, 
for each player $i$, we can define a partial order 
$\below_i$.  
on  player $i$'s information sets such that $I \below_i I'$ if,  
for every history $h \in I$, there is a prefix $h'$ of $h$ in
$I'$. Thus, $I\below_i I'$ if $I$ is below~$I'$ in the game tree.  
We define the {\em height}  of an information set~$I$ for player $i$,
denoted by $\height(I)$, inductively as follows; 
$\height(I)=1$ if $I$
is a maximal set for player $i$, that is, there is no information set
$I'$ such that $I' \below_i I$; and if $I$ is not maximal, then   
$\height(I)=\max\{\height(\hat{I})+1: \hat{I}\below_i I\}$. Since $\Gamma$
is a finite game, $\height(I)$ is well defined. Indeed, the size of the
game ensures that there is a finite bound~$d$ such that $\height(I)\le
d$ for all information sets in the game. 
}

We are now ready to formally capture perfect equilibrium in terms of
common knowledge of rationality,  
using Theorem~\ref{thm:thpe}. Intuitively, the assumption that
$\sigma_i$ is a best response relative to the nonstandard $\vecsig'$ 
is replaced by the assumption of common knowledge of rationality when
players play $\vecsig'$. 
%

\thm\label{cor:thpe} 
Let $\Gamma$ be a finite extensive-form game with perfect recall. Then 
$\vec{\sigma}$ is a perfect equilibrium of $\Gamma$ iff there
exist a
normal non-Archimedean field $\IR^*$,
a
nonstandard, 
completely-mixed
strategy profile $\vecsig'$ that differs infinitesimally from
$\vec{\sigma}$ with probabilities in $\IR^*$,
and a 
model $M = (\Omega,\hist, (\Pr_i)_{i \in N})$ of
$\Gamma$ 
compatible with $\vecsig'$ 
where local rationality is common knowledge.
\ethm

\prf Suppose that $\vec{\sigma}$ is a perfect equilibrium of $\Gamma$.
Then, by Theorem~\ref{thm:thpe}, 
there exists a normal non-Archimedean field $\IR^*$ and a 
nonstandard
completely-mixed strategy profile~$\vec{\sigma}'$ 
with probabilities in $\IR^*$
that differs
infinitesimally from $\vec{\sigma}$ such that, for each player~$i$, 
the strategy~$\sigma_i$ is a local best 
response for~$i$ relative to $\vec{\sigma}'$.
Let $M = (\Omega,\hist,(\Pr_i)_{i \in N})$ be such that $\Omega =
\{\omega_{h}: h \in Z_\Gamma\}$, $\hist(\omega_h) =h$, and
$\Pr_i(\omega_{h}) = \Pr_{\vec{\sigma}'}(h)$, for $i = 1,\ldots, n$.
Clearly  $M$ is compatible with $\vecsig'$.  
We claim that it is common knowledge in $M$ that all players are
locally rational. 

To see this, consider an arbitrary state 
$\omega_h\in\Omega$.
Suppose that 
$I$ is an information set for player $i$,
$h'\in I$ is a prefix of~$h$, the action played  by~$i$ at~$h'$ in~$h$
is~$a$, and  
$\stand{\sigma'_i(I))(a)} > 0$.  
Since $\sigma_i$ is a local best response for $i$ conditional on having reached $I$
using $\vecsig'$, 
Equation~(\ref{eq2})  from Definition~\ref{def:local-best}
(with~$\eps=0$)  implies 
that
{\red{
$$
\EU_i((\sigma'_i[I/\sigma_i(I)],\vecsig'_{-i}),\mu_I^{\vecsig'})\mid I) \ge
\EU_i((\sigma_i[I/a'],\vecsig'_{-i}),\mu_I^{\vecsig'})\mid I)
$$
for all $a' \in \Delta(A_I)$.  
It easily follows that 
\begin{equation}\label{eq3}
\EU_i((\sigma'_i[I/a''],\vecsig'_{-i}),\mu_I^{\vecsig'})\mid I) \ge
\EU_i((\sigma_i[I/a'],\vecsig'_{-i}),\mu_I^{\vecsig'})\mid I)
\end{equation}
for all actions~$a' \in A_I$ and 
all actions $a''$ in the support of $\sigma_i(I)$.
By assumption, $\sigma_i'$ differs infinitesimally
from~$\sigma_i$. Hence, the  
fact that $\stand{\sigma'_i(I)(a)} > 0$ implies that $\sigma_i(I)(a) >
0$, so that the  
action $a$ must be in the support of $\sigma_i(I)$.
Therefore,  (\ref{eq3}) holds 
for $a'=a$, so~$i$ is
rational at~$\omega_h$. 
We conclude that every player~$i$ is locally rational at all states
$\omega\in\Omega$ and thus, by definition,  
it is common knowledge in~$M$ that the players are locally rational.
}}

For the converse, 
fix~$\vecsig$ and
suppose that  there 
exist $\IR^*$, $\vecsig'$, 
and 
a model $M$ as required by
the theorem.  
\sout{
We now show by induction on $\height(I)$ 
%
that, at each information set~$I$
for player~$i$, 
the strategy
$\sigma_i$ is a best response to $\vecsig'_{-i}$
conditional on having reached $I$ using $\vecsig'$.  The 
claim that $\vecsig$ is a perfect equilibrium in~$\Gamma$ 
then follows from Theorem~\ref{thm:thpe}.
\ldots [Details of the argument omitted, but moved to proof of next theorem.]}
\commentout{
Note that it easily follows
from the assumptions that $\sigma_i$ is ``locally'' a best
response; that is, we cannot just make a change at information set $I$
that will improve $i$'s utility.   We must show that $\sigma_i$
continues to be a best response even if we allow changes at $I$ and all
information sets for $i$ below $I$
in the game tree.
If $\height(i)=1$ then 
$I$ is one of the maximal
information sets in the partial order
and 
there are no information sets for~$i$ below~$I$, so this is trivially true.   
Now suppose that 
$\height(I)=k>1$ 
and the result holds for all
information sets of height~$<k$.
Further suppose, by way of
contradiction, that there 
exists a strategy $\tau_i$ for $i$ such that 
\begin{equation}
\label{tau-contradiction}
\EU_i(((\tau_i,\vec{\sigma}'_{-i}),\mu^{\vecsig'}_I) \mid I) >
\EU_i(((\sigma_i,\vec{\sigma}'_{-i}),\mu^{\vecsig'}_I) \mid I).
\end{equation}
There must be some action $a'$ in the support of $\tau_i(I)$ and 
an action $a$ in the support of $\sigma_i(I)$ such that 
$$\EU_i(((\tau_i[I/a'],\vec{\sigma}'_{-i}),\mu^{\vecsig'}_I) \mid I) >
\EU_i(((\sigma_i[I/a],\vec{\sigma}'_{-i}),\mu^{\vecsig'}_I) \mid I).$$
Let $\setI=\{I_1, \ldots, I_m\}$ be the 
information sets for player $i$ that immediately succeed~$I$ 
in~$\Gamma$ (i.e., for each $I_j\in\setI$, it is not the case that there is an
information set 
$I'$ such that $I_j \below_i I' \below_i I$)
and can be reached by starting at a history in $I$ and
playing $(\tau_i(I),\vec{\sigma}'_{-i})$.
the inductive hypothesis guarantees that 
$\sigma_i$ is a best response to $\vecsig'_{-i}$ at each
information set $I'$ for $i$ below $I$, player $i$'s utility is at least
as high if he plays $\sigma_i$ rather than $\tau_i$ at 
each $I'\in\setI$.
Thus,
$$\EU_i(((\sigma_i[I/a'],\vec{\sigma}'_{-i}),\mu^{\vecsig'}_I) \mid I)
\ge \EU_i(((\tau_i[I/a'],\vec{\sigma}'_{-i}),\mu^{\vecsig'}_I) \mid I),$$ 
and hence 
\begin{equation}
\label{two star}
(((\sigma_i[I/a'],\vec{\sigma}'_{-i}),\mu^{\vecsig'}_I) \mid I)
> \EU_i(((\sigma_i[I/a],\vec{\sigma}'_{-i}),\mu^{\vecsig'}_I) \mid I). 
 \end{equation}

Let history $h$ be a history that goes through $I$ such that $i$ plays
action $a$ at $I$ in history $h$.  Since $a$ is in the support of
$\sigma_i$, it must be the case that $\stand{\sigma'_i(I)(a)} > 0$.  
By (\ref{two star}),
it now follows that player~$i$ is not rational at $\omega_h$,
contradicting the assumption  
that rationality of all players is common knowledge in~$M$.
}
{\red{For each information set $I$ for player $i$, if 
$a \in A_I$ is in
  the support of $\sigma_i(I)$, then $\stand{\sigma_i'(I)(a))} >0$.
Since $M$ is compatible with $\vecsig'$, there must exist some state~$\omega$ in $M$ with a prefix $h$ of $\hist(\omega)$ in $I$ such that
$i$ plays $a$ after $h$ in $\hist(\omega)$.  Since $i$ is
locally
rational at $\omega$, 
performing 
$a$ must 
be  a local best response for $i$ conditional on having reached $I$
using $\vec{\sigma}'$.
Thus, $\sigma_i(I)$ must be a local best response for $i$ conditional
on having reached $I$ using $\vecsig'$.
Hence, by 
Theorem~\ref{thm:thpe}
we obtain that 
$\vec{\sigma}$ is a perfect
equilibrium.}}
\eprf

Perhaps not surprisingly, we 
obtain 
an analogue of Theorem~\ref{cor:thpe}
by replacing ``local rationality'' by ``rationality''.
\thm\label{cor:quasi} 
Let $\Gamma$ be a finite extensive-form game with perfect recall. Then 
$\vec{\sigma}$ is a quasi-perfect equilibrium of $\Gamma$ iff there
exist a 
normal non-Archimedean field $\IR^*$,
a nonstandard, 
completely-mixed
strategy profile $\vecsig'$ that differs infinitesimally from
$\vec{\sigma}$ with probabilities in $\IR^*$,
and a 
model $M = (\Omega,\hist, (\Pr_i)_{i \in N})$ of
$\Gamma$ compatible with $\vecsig'$ 
where rationality is common knowledge.
\ethm

\prf  The proof is similar in spirit to that of
Theorem~\ref{cor:thpe}, and simpler, so we leave details to the
reader.
\eprf

Interestingly, for sequential equilibrium, we can work 
with either $\eps$-rationality or $\eps$-local rationality.
\commentout{
To see this, consider a state $\omega_{\vecS} \in\Omega$ 
and a player~$i$. 
By assumption, $\sigma_i$ is a best response for~$i$ relative to
$\vec{\sigma}'$. 
That is, for all strategies $S_i' \in \S_i$ 
and information sets~$I$ for~$i$, 
we have that 
\[
{\Pr}_{i}^{\vecsig'}([\sigma'_i,I,\sigma_i],\vecsig'_{-i})\ge {\Pr}_{i}^{\vecsig'}([\sigma'_i,I,S_i],\vecsig'_{-i},)
\]
and thus 
$$\sum_{h \in H} (\sigma_i,\vec{\sigma}'_{-i})(h \mid I) u_i(h) \ge 
\sum_{h \in H} (S_i,\vec{\sigma}'_{-i})(h \mid I) u_i(h).$$

{\bf ZZZZ changed to here}

Given how $\Pr_i$ is defined,
it is easy to check that, for any strategy $S_i' \in \S_i$ such that 
$(S_i',\vec{\sigma}_{-i})$ reaches $I$ with positive probability, we
have 
\begin{equation}\label{eq1}
\sum_{h \in H} (S_i',\vec{\sigma}'_{-i})(h \mid I) u_i(h) = 
\sum_{\vecS_{-i}' \in \S_{-i}}
\Pr_i^{I,S_i}([\vecS_{-i}'])u_i(\vecS').
\end{equation}
Here are the details:
$$\begin{array}{ll}
&\sum_{h \in H} (S_i',\vec{\sigma}'_{-i})(h \mid I) u_i(h) \\
= &\sum_{\vecS_{-i}' \in \S_{-i}}
(S_i',\vec{\sigma}'_{-i})((\vecS') \mid I) u_i(\vecS')\\
= &\sum_{\vecS_{-i}' \in \S_{-i}} \vec{\sigma}'_{-i}(\vecS'_{-i} \mid I)
u_i(\vecS')\\ 
= &\sum_{\vecS_{-i}' \in \S_{-i}} \vec{\sigma}(\vecS'_{-i} \mid I)
u_i(\vecS')\\ 
= &\sum_{\vecS_{-i}' \in \S_{-i}} \Pr_i([\vecS'_{-i}] \mid [I])
u_i(\vecS')\\ 
= &\sum_{\vecS_{-i}' \in \S_{-i}} \Pr_i^{I,S_i}(\vecS'_{-i})
u_i(\vecS').
\end{array}$$

It immediately follows that $i$ is rational at $\omega_{\vecS}$
conditional on reaching~$I$.


normal non-Archimedean field $\IR^*$, 
a
nonstandard, completely-mixed
strategy profile $\vecsig'$ that differs infinitesimally from
$\vec{\sigma}$ with probabilities in $\IR^*$,
and a model $M = (\Omega,\strat, (\Pr_i)_{i \in N})$ of $\Gamma$ 
such that $\Pr_i$ is a nonstandard probability measure with values in
$\IR^*$ for $i \in N$ and it is 
almost 
common knowledge in~$M$ that the 
players are always rational
and  $\Pr_i^{\S} = \Pr_{\vecsig'}$ for all $i \in N$.
We claim that, for each player $i$ and information set $I$ for player
$i$, $\sigma_i$ is a best response to $\vec{\sigma}_{-i}'$, conditional
on reaching $I$.  To see this, it suffices to show that for each player
$i$, strategy $S_i$ in the support of $\sigma_i$, and information set
$I$, $S_i$ is a best response to $\vec{\sigma}_{-i}'$ conditional on
reaching~$I$.   

Fix $I$.  We can assume without loss of generality that
$(S_i,\vec{\sigma}_{-i}')(I) > 0$, or there is nothing to prove.
Choose $\vecS_{-i}$ such that $(S_i, \vecS_{-i})$ 
is
in the
support of $\vec{\sigma}$.  
Since $\vec{\sigma}'$ differs infinitesimally from
$\vec{\sigma}$ and $\Pr_i = \Pr_{\vec{\sigma}'}$, it must be the case
that $\stand{\omega_{\vec{S}}} > 0$.  Thus, $i$ is rational 
conditional on reaching $I$ at state $\omega_{\vec{S}}$.  
As above, we can show that (\ref{eq1}) holds, so $S_i$ must be a best
response to $\vec{\sigma}_{-i}$ conditional on reach $I$.  It now
follows from Theorem~\ref{cor:thpe} that $\vec{\sigma}$ is a perfect
equilibrium.  
\eprf
}

\thm\label{cor:seqeq} 
Let $\Gamma$ be a finite extensive-form game with perfect recall. 
The following are equivalent:
\begin{itemize}
\item[(a)]
there exists a belief system $\mu$ such that the 
assessment $(\vec{\sigma},\mu)$ is a sequential equilibrium of
$\Gamma$;
\item[(b)] there
exist a normal non-Archimedean field $\IR^*$,
a nonstandard,
completely-mixed
behavioral
strategy profile $\vecsig'$ 
with probabilities in $\IR^*$ 
that differs infinitesimally from~$\vec{\sigma}$, an infinitesimal 
$\eps > 0$ in~$\IR^*$, and 
a model $M = (\Omega,\hist, (\Pr_i)_{i \in N})$ compatible with
$\vecsig$ 
where $\eps$-rationality is common knowledge; 
\item[(c)]
there exist a normal non-Archimedean field $\IR^*$,
a nonstandard, completely-mixed
strategy profile $\vecsig'$ 
with probabilities in $\IR^*$ 
that differs infinitesimally from
$\vec{\sigma}$, an infinitesimal 
$\eps > 0$ in~$\IR^*$, and 
a model $M = (\Omega,\hist, (\Pr_i)_{i \in N})$ compatible with
$\vecsig$ 
where $\eps$-local rationality is common knowledge. 
\end{itemize}
\ethm

\prf 
In light of Theorem~\ref{thm:seqeq}, the equivalence of (a) and (b) is
almost immediate.
\sout{
The proof is 
similar 
to that of Theorem~\ref{cor:thpe},
using Theorem~\ref{thm:seqeq} in place of Theorem~\ref{thm:thpe}.}
To see that (a) implies (c), suppose that
$(\vecsig,\mu)$ is a sequential equilibrium.  By
    Theorem~\ref{thm:seqeq}, there exists a strategy profile
    $\vecsig'$ that differs infinitesimally from $\vecsig$ and an
    infinitesimal~$\eps$ such that, for each player $i$, 
    strategy
    $\sigma_i$ is
    an $\eps$-local best response relative to $\vecsig'_{-i}$.  
Construct $M$ as in the proof of Theorem~\ref{cor:thpe}. 
Since $\sigma_i$ differs infinitesimally from $\sigma'_i$, for each
player $i$, there
exists an infinitesimal $\eps'_i$ such that, for all information
sets $I$ for player $i$,
\begin{equation}\label{eq3.5}
\EU_i(((\sigma_i,\vec{\sigma}'_{-i}),\mu^{\vecsig'}_I) \mid I)  \ge
\EU_i(((\sigma_i,\vec{\sigma}'_{-i}),\mu^{\vecsig'}_I) \mid I) - \eps'_i.
\end{equation}
Let $\eps' = \max_{i = 1,\ldots, n} \eps'_i$ and let 
\[r = \min\nolimits_{i = 1,\ldots, n}\{\sigma_i(I)(a):~\mbox{$I$ is an information  set  for $i$ and $a$ is in the support of $\sigma_i(I)\}$};\]
that is, $r$ is the smallest positive probability
assigned by a strategy $\sigma_i$, $i=1,\ldots,n$.  Note that $\eps' +
\eps + \eps/r$ is an infinitesimal (since $r$ is a standard rational).
We claim that $(\eps' + \eps + \eps/r)$-local rationality is common
knowledge in $M$.   

To see this, 
fix a player $i$ and a state $\omega$ in $M$, and let $h=\hist(\omega)$.  
Again, suppose that~$I$ is an information set for player $i$,
$h'\in I$ is a prefix of~$h$, the action played  by~$i$ at~$h'$ in~$h$
is~$a$, and  $\stand{\sigma'_i(I))(a)} > 0$.   We want to show that
\begin{equation}\label{eq4}
\EU_i(((\sigma_i'[I/a],\vec{\sigma}'_{-i}),\mu^{\vecsig'}_I) \mid I) \ge
\EU_i(((\sigma_i'[I/a'],\vec{\sigma}'_{-i}),\mu^{\vecsig'}_I) \mid
I) - (\eps' + \eps + \eps/r)
\end{equation}
for all actions $a' \in A_I$.  
First observe that, by 
choice of $\eps'$, it easily follows 
from (\ref{eq3.5})
that 
\begin{equation}\label{eq5}
\EU_i(((\sigma_i'[I/a],\vec{\sigma}'_{-i}),\mu^{\vecsig'}_I \mid I)  \ge
\EU_i(((\sigma_i[I/a],\vec{\sigma}'_{-i}),\mu^{\vecsig'}_I \mid I) -
\eps'.
\end{equation}
Moreover, since 
 $\sigma_i$ is an $\eps$-best response relative to
$\vecsig'_{-i}$, for all actions  $a' \in A_I$, we must have 
\begin{equation}\label{eq7}
\EU_i(((\sigma_i,\vec{\sigma}'_{-i}),\mu^{\vecsig'}_I) \mid I) \ge
\EU_i(((\sigma_i'[I/a'],\vec{\sigma}'_{-i}),\mu^{\vecsig'}_I) \mid
I) - \eps.\end{equation}
Since (\ref{eq7}) holds for each action $a'$ in the support of
$\sigma_i(I)$, we must have
$$\begin{array}{ll}
& \EU_i(((\sigma_i,\vec{\sigma}'_{-i}),\mu^{\vecsig'}_I) \mid I)  \\ 
=
&\sigma_i(I)(a)\EU_i(((\sigma_i[I/a],\vec{\sigma}'_{-i}),\mu^{\vecsig'}_I)
\mid I)  + 
\sum_{\{a': \, \sigma_i(I)(a') > 0,\, a' \ne a\}}
\sigma_i(I)(a')\EU_i(((\sigma_i[I/a'],\vec{\sigma}'_{-i}),\mu^{\vecsig'}_I)
\mid I)  \\
\le &\sigma_i(I)(a)\EU_i(((\sigma_i[I/a],\vec{\sigma}'_{-i}),\mu^{\vecsig'}_I)
\mid I)  +
\sum_{\{a': \, \sigma_i(I)(a') > 0,\, a' \ne a\}}
\sigma_i(I)(a')(\EU_i(((\sigma_i,\vec{\sigma}'_{-i}),\mu^{\vecsig'}_I)
+ \eps)\\
= &\sigma_i(I)(a)\EU_i(((\sigma_i[I/a],\vec{\sigma}'_{-i}),\mu^{\vecsig'}_I)
\mid I)  + 
(1-\sigma_i(I)(a))(\EU_i(((\sigma_i,\vec{\sigma}'_{-i}),\mu^{\vecsig'}_I)
\mid I) + \eps).\end{array}$$
A little algebraic manipulation now shows that 
\begin{equation}\label{eq6}
\begin{array}{lll}
\EU_i(((\sigma_i[I/a],\vec{\sigma}'_{-i}),\mu^{\vecsig'}_I)
\mid I) &\ge &\EU_i((\sigma_i,\vec{\sigma}'_{-i}),\mu^{\vecsig'}_I) \mid
I) - 
\eps(1-\sigma_i(I)(a))/\sigma_i(I)(a)\\
&\ge &\EU_i((\sigma_i,\vec{\sigma}'_{-i}),\mu^{\vecsig'}_I) \mid
I) - \eps/r.
\end{array}
\end{equation}
Equation
(\ref{eq4}) follows immediately from 
(\ref{eq5}), (\ref{eq7}),  and (\ref{eq6}).  
Thus, $(\eps' + \eps + \eps/r)$-local rationality is common knowledge in $M$.  
We have shown that (a) implies (c).  
\sout{For the converse, the argument is similar to that in the
previous proof, but it involves an additional subtlety.  }

\red{
It remains to show that (c) implies 
(a).
So suppose that there exists a field $\IR^*$, a nonstandard strategy
profile $\vecsig'$, an infinitesimal $\eps > 0$ in $\IR^*$, and a 
model $M$ where $\eps$-local rationality is common knowledge, as
required for (c) to hold.  It is
almost immediate that $\sigma_i$ is an $\eps$-local best response
for $i$ relative to $\sigma'_{_i}$.  We want to show that there exists
some infinitesimal $\eps''$, 
possibly different from $\eps$, such that
$\sigma_i$ is an $\eps''$-best response for $i$ relative to
$\sigma'_i$, for each player $i$.  The result then follows from
Theorem~\ref{thm:seqeq}.  

To do this, we need some preliminary definitions.
In a finite extensive-form game~$\Gamma$ with perfect recall, 
for each player $i$, we can define a partial order $\below_i$
on  player $i$'s information sets such that $I \below_i I'$ if,  
for every history $h \in I$, there is a prefix $h'$ of $h$ in~$I'$. 
Thus, $I\below_i I'$ if $I$ is below
(i.e., appears later than)
$I'$ in the game tree.  
We define the {\em height}  of an information set~$I$ for player $i$,
denoted by $\height(I)$, inductively as follows; 
$\height(I)=1$ if $I$
is a maximal set for player $i$, that is, there is no information set
$I'$ such that 
$I' \below_i I$. If~$I$ 
is not maximal, then   
$\height(I)=\max\{\height(\hat{I})+1: \hat{I}\below_i I\}$. Since $\Gamma$
is a finite game, $\height(I)$ is well defined. Indeed, the size of the
game ensures that there is a finite bound~$d$ such that $\height(I)\le
d$ for all information sets in the game.   
For $\eps'$ defined just before Equation~(\ref{eq3.5}), 
we can now prove the following 
result:
}\black{}

\lem\label{localtoglobal} 
$\sigma_i$ is a $d(\eps + \eps')$-best
response for $i$ relative to $\vecsig'$ in $\Gamma$.
\elem

\prf
For all
information sets~$I$ of player~$i$, we 
show by induction on $k=\height(I)$ that $\sigma_i$ is a {\red{$k(\eps + \eps')$-best}}
response to $\vecsig'_{i}$ conditional on 
having reached $I$ using $\vecsig'$. 
So fix an arbitrary player~$i$, and
let~$I$ be an information set for~$i$.  
If $I$ is maximal, 
then $\height(I)=1$. 
By assumption, 
$\sigma_i$ is a {\red{local}} $\eps$-best response to $\vec{\sigma}'_{-i}$
conditional on having 
reached $I$ using $\vecsig'$, so the base case of the induction holds.
Now suppose that $\height(I)=k>1$ and 
that the claim holds for all~$I'$ such that 
$\height(I') < k$.
\sout{The argument closely follows the one in the previous proof. 
Rather than (\ref{tau-contradiction}), we now.}

{\red{
By choice of $\eps'$, we have 
by Equation (\ref{eq3.5}) 
that
\begin{equation}\label{eq9}
\EU_i(((\sigma_i,\vec{\sigma}'_{-i}),\mu^{\vecsig'}_I) \mid I) \ge
\EU_i(((\sigma'_i[I/\sigma_i(I)],\vec{\sigma}'_{-i}),\mu^{\vecsig'}_I) \mid I) 
 - \eps'.
\end{equation}
Let $\tau_i$ be an arbitrary strategy for player $i$.  
By assumption, $\sigma_i$ is a local $\eps$-best response relative to
$\vecsig_{-i}'$, so 
\begin{equation}\label{eq8}
\EU_i(((\sigma'_i[I/\sigma_i(I)],\vec{\sigma}'_{-i}),\mu^{\vecsig'}_I) \mid I) 
\ge \EU_i(((\sigma'_i[I/\tau_i(I)],\vec{\sigma}'_{-i}),\mu^{\vecsig'}_I) \mid
I) - \eps
.
\end{equation}
Let $\setI=\{I_1, \ldots, I_m\}$ be the 
information sets for player $i$ that immediately succeed~$I$ 
in~$\Gamma$ (i.e., for each $I_j\in\setI$, $I_j \succeq I$ and 
there is no
information set~$I'$ such that $I_j \below_i I' \below_i I$)
and can be reached by starting at a history in $I$ and
playing $\tau(I)$.}}
By the inductive hypothesis,
$\sigma_i$ is a $(k-1)(\eps+\eps')$-best response to $\vecsig'_{-i}$ at each
information set $I'\in\setI$, so player $i$'s utility is at most
$(k-1)(\eps+\eps')$ worse  
if he plays $\sigma_i$ rather than $\tau_i$ at 
each $I'\in\setI$.  It easily follows that
\begin{equation}\label{eq10}
\EU_i((\sigma_i[I/\tau_i(I)],\vecsig'_{-i}),\mu_I^{\vecsig})\mid I) \ge
\EU_i((\tau_i,\vecsig'_{-i}),\mu_I^{\vecsig})\mid I) - (k-1)(\eps+\eps').
\end{equation}
Putting together (\ref{eq9}), (\ref{eq8}), and (\ref{eq10}), we 
obtain
that 
$$\EU_i(((\sigma_i,\vec{\sigma}'_{-i}),\mu^{\vecsig'}_I)
\mid I)  \ge \EU_i(((\tau_i,\vec{\sigma}'_{-i}),\mu^{\vecsig'}_I) \mid
I)
 - k(\eps + \eps').
$$

\commentout{
Putting together the pieces, we have
$$
\begin{array}{ll}
&\EU_i(((\sigma_i,\vec{\sigma}'_{-i}),\mu^{\vecsig'}_I) \mid I) \\
= &\EU_i((\sigma_i[I/a],\vecsig'_{-i}),\mu_I^{\vecsig})\mid I) \\
\ge &\EU_i((\sigma_i[I/a'],\vecsig'_{-i}),\mu_I^{\vecsig})\mid I) - \eps \\
\ge &\EU_i(((\tau_i[I/a'],\vec{\sigma}'_{-i}),\mu^{\vecsig'}_I) \mid I) -
(k-1)\eps -\eps \\
= &\EU_i(((\tau_i[I/a'],\vec{\sigma}'_{-i}),\mu^{\vecsig'}_I) \mid I) - k\eps, 
\end{array}$$
contradicting (\ref{tau-contradiction}). 
}

Since $\height(I) \le d$ for each information set $I$ in $\Gamma$, it
follows that  $\sigma_i$ is an 
$d(\eps+\eps')$-best response for $i$ relative to $\vecsig'$,  
for each player $i=1,\ldots,n$.  This completes the proof of the
lemma.
\eprf

Clearly $\eps''=d(\eps + \eps')$ is an infinitesimal, so by
Theorem~\ref{thm:seqeq}, it  follows that  
there exists a belief system $\mu$ such that the assessment
$(\vecsig,\mu)$ is a sequential equilibrium of $\Gamma$,
as desired.
\eprf

It follows from Theorem~\ref{cor:seqeq} that Theorem~\ref{thm:seqeq}
can be generalized to use either $\eps$-rationality or $\eps$-local
rationality.  Each of the  choices gives
a characterization of sequential rationality.  

It is interesting to compare our results to those of Asheim and Perea
\citeyear{AP05}.   As 
mentioned,
they provide epistemic characterizations of
sequential equilibrium and quasi-perfect equilibrium in 2-player games
in terms of rationality.  Their notion of rationality
is essentially equivalent to ours; since they do not use local
rationality, it is perhaps not surprising that they do not deal with
perfect equilibrium, which seems to require it.  

To obtain their results, Asheim and Perea
represent uncertainty using a generalization of LPSs (lexicographic probability
sequences) \cite{BBD1,BBD2} that they call \emph{systems of
  conditional lexicographic probabilities} (SCLPs).  An LPS is a
sequence $(\Pr_0, \ldots, \Pr_k)$ of probability measures on a measure
space $(S,\F)$.  Roughly
speaking, we can identify such a sequence with the nonstandard
probability measure $(1-\epsilon - \cdots - \epsilon^k)\Pr_0 +
\epsilon\Pr_1 + \cdots + \epsilon^k \Pr_k$ on $(S,\F)$.  Indeed, 
it has been shown that 
LPSs and
nonstandard probability spaces (NPSs) are
essentially equivalent in finite spaces \cite{BBD1,Hal26}.
However, it is not hard to
show that SCLPs can capture some situations that cannot be captured by
NPSs.  Roughly speaking, this is because SCLPs do not necessarily
satisfy  an analogue of the chain rule of probability ($\Pr(A\mid B)
\times \Pr(B\mid C) = \Pr(A\mid C)$ if $A \subseteq B \subseteq C$),
which does hold for NPSs.%
\footnote{There is no notion of multiplication in SCLPs, so 
this statement
  is not quite accurate.  Nevertheless, consequences of the chain
  rule, such as 
  that 
  $\mu(A \mid B) = \mu(A' \mid B)$ implies $\mu(A \mid
  C) = \mu(A' \mid C)$ do not hold for SCLPs.}
(Of course, we might view such situations as unreasonable.)
It would be interesting to investigate whether our results could be
obtained with some variant of LPSs or CPSs (conditional probability
spaces).  

Another relatively minor difference between our result and that of Asheim and
Perea 
is that they
work with what they call \emph{common certain belief} rather
than 
with
common knowledge, where 
certain belief of $E$ 
is defined
relative to a
model characterized by an LPS $(\mu_1, \ldots, \mu_k)$ if $\mu_j(E) =
1$ for $j = 1, \ldots, k$.
Although Asheim and Perea's theorems are stated in terms of
  \emph{mutual} certain belief of rationality rather than common certain belief,
  where mutual certain belief holds 
  if
  both of the players have
  certain belief of rationality, they also require mutual certain belief
  of each player's type; in their setting, this implies common certain
  belief of rationality.

Finally, in their characterization of quasi-perfect equilibrium,
Asheim and 
Perea 
also require common certain belief of
\emph{caution}, which, roughly speaking, in our language, says that
players should prefer a strategy that is a best response to one that
is an $\eps$-best response, even for an infinitesimal $\eps$.
Dropping caution when moving from quasi-perfect equilibrium to
sequential equilibrium in Asheim and Perea's framework corresponds
to moving from rationality to $\eps$-rationality in our framework.

\section{Discussion}\label{sec:discussion}
Theorems~\ref{cor:thpe}, \ref{cor:quasi}, and \ref{cor:seqeq}
illustrate the role 
that common knowledge of rationality plays in perfect
equilibrium, quasi-perfect equilibrium, and sequential equilibrium.  Comparing
Theorem~\ref{thm:charNE} to Theorem~\ref{cor:thpe}, note that for
$\vecsig$ to be a perfect equilibrium, 
Theorem~\ref{cor:thpe} requires players to \emph{always} be rational;
that is, 
for every information set~$I$ that a player $i$ can reach in the game, $i$
must be rational conditional on reaching~$I$. 
Since Theorem~\ref{thm:charNE} considers only normal-form games, the
requirement that players always be rational has no bite.  But we could
prove an analogue of Theorem~\ref{thm:charNE} for Nash equilibrium in
extensive-form games, and again it would suffice to have rationality
\emph{ex ante}, rather than conditional on reaching each information
set.  The other key difference between Theorems~\ref{thm:charNE}
and~\ref{cor:thpe} is that in Theorem~\ref{cor:thpe}, 
rather than taking the probability on histories in $M$
to be determined by $\vecsig$, it is determined by $\vecsig'$, a
completely-mixed nonstandard strategy that differs infinitesimally from
$\vecsig$.  
Note that there are many strategies
that differ infinitesimally from $\vecsig$.  The exact choice of $\vecsig'$ 
has only an infinitesimal impact on $i$'s beliefs at information sets $I$
that are 
on the equilibrium path; but for information sets $I$ off the
equilibrium path, the choice of 
$\vecsig'$ completely determines $i$'s beliefs; different choices can
result in quite different beliefs.

The 
distinction
between 
Theorems~\ref{cor:thpe} and~\ref{cor:seqeq} highlights 
one way of thinking about the difference between perfect equilibrium and
sequential equilibrium.  For perfect equilibrium, it has to be common
knowledge that players are 
always rational; for sequential equilibrium, it
suffices 
to have common knowledge that players are always
$\eps$-rational for an infinitesimal
$\eps > 0$.
The distinction between Theorems~\ref{cor:thpe} and~\ref{cor:quasi}
brings out the point that van Damme already stressed in the definition
of quasi-perfect equilibrium: the difference
between local best responses and best responses.  We find it of
interest that this distinction does not play a role in sequential
equilibrium. 

Our results complement Aumann's earlier epistemic characterizations of Nash
and of correlated equilibria. The general picture obtained is that all of 
these solution concepts can be characterized in terms of
common knowledge of rationality; the differences 
between the characterizations 
depend on what we
assume about the prior probability, whether rationality holds at all
information sets or just at the beginning, and whether we consider
rationality or $\eps$-rationality.  As we show in 
related work
\cite{HM10}, as a consequence of this observation, it follows that all
these solution concepts can be embodied in terms of a single
\emph{knowledge-based program} \shortcite{FHMV,FHMV94}, which essentially
says that player~$i$ should perform 
action~$a$ 
if she believes 
both 
that she plans to perform~$a$ and that 
playing~$a$ is optimal for her in the sense of being a best
response. 
This is, arguably, the essence of rationality.  
In the case of each of the equilibrium notions that we have discussed, 
for the corresponding notions of rationality and best response, if it 
is common knowledge that everyone is following 
this
knowledge-based program, then rationality is common knowledge.

Can other standard solution concepts be characterized this way?  
It is 
straightforward 
to state and prove an analogue of
Theorem~\ref{thm:charNE} for Bayes-Nash equilibrium.  Now the state
space in the model would include each 
player's
type.  If we define
rationality and best responses in terms of minimax regret, rather than
in terms of maximizing expected utility, Boutilier and Hyafil
\citeyear{BH04} define a notion of \emph{minimax-regret} equilibrium
that can be captured in terms of common knowledge of rationality.
Similarly, Aghassi and Bertsimas \citeyear{AB06} define rationality in
terms of maximin (i.e., maximizing the worst-case utility) and use that
to define what they call \emph{maximin equilibria}.  Again, we can prove
an analogue of Theorem~\ref{thm:charNE} for this solution concept.

Perhaps more interesting is the solution concept of \emph{iterated
admissibility}, also known as \emph{iterated deletion of weakly
dominated strategies}.  Brandenburger, Friedenberg, and Keisler
\citeyear{BFK04}
provide an epistemic characterization of iterated admissibility (i.e.,
iterated deletion of weakly dominated strategies) where uncertainty is
represented using LPSs (lexicographic probability sequences). 
They define a notion of belief (which they call
\emph{assumption}) appropriate for their setting, and show that
strategies that survive $k$ rounds of iterated deletion are ones that
are played in states where 
there is $k$th-order mutual belief in
rationality; that is, everyone assumes that everyone assumes \ldots ($k-1$
times) that everyone is rational.  However, they prove only that their
characterization of iterated admissibility holds in particularly rich
structures called \emph{complete} structures, where all types are 
possible.  However, more recently, Halpern and Pass \citeyear{HP11a}
provide a characterization that is closer to the spirit of
Theorem~\ref{thm:charNE}.    The key new feature is that instead of just
requiring that everyone is rational, and that everyone knows 
that
everyone is
rational, and that everyone knows that everyone knows \ldots, they
require that 
\emph{all} everyone knows is that everyone is rational, and
that all everyone knows is that all everyone knows is 
that
everyone is
rational, and so on.  
In this claim, the statement that \emph{all agent $i$ knows is $\phi$} 
is true at a state
$\omega$ if, not only is it the case that $\phi$ 
is true at all states
that $i$ considers possible at $\omega$ (which is what is required for
$i$ to know $\phi$ at $\omega$), but 
it is also the case 
that $i$ assigns $\psi$ positive
probability for each formula $\psi$ consistent with $\phi$.  Thus, we
capture ``all $i$ knows is $\phi$'' by requiring that $i$
considers any situation compatible with $\phi$ possible.  In the
specific case of iterated admissibility, this means that~$i$ considers
possible (i.e., assigns positive probability to) all strategies
compatible with rationality.  As shown by Halpern and Pass
\citeyear{HP11a}, a strategy 
survives
$k$ rounds of iterated
deletion iff it is played at a state in a structure where all everyone
knows is that all everyone knows \ldots ($k$ times) that everyone is
rational.  This result does not require the restriction to complete
structures.

Now consider 
\emph{extensive-form rationalizability} (EFR)
\cite{Pearce84}, an extension of  rationalizability that seems
appropriate for extensive-form games \cite{HP11a}.  Battigalli and
Siniscalchi \citeyear{BS02} provide an epistemic characterization of EFR
using a notion of \emph{strong belief}; these are beliefs that are
maintained unless evidence shows that the beliefs are inconsistent. 
For example, if player 1 has a strong belief of player 2's rationality,
then whatever moves player 2 makes, player 1 will revise her beliefs
and, in particular, her beliefs about player 2's beliefs, in such a way
that she continues to believe that player 2 is rational (so that she
believes that player 2 is making a best response to his beliefs), unless
it is inconsistent for her to believe that player 2 is rational.
Battigalli and Siniscalchi characterize EFR in terms of
common strong belief of rationality.  
Specifically, 
they show that a
strategy satisfies EFR iff it is played in a complete structure.  Again,
using  ``all $i$ knows'' 
would
allow us to give an epistemic characterization
of EFR in the spirit of the theorems in this paper
without  the restriction to complete structures  
\cite{HP11a}.%
\footnote{Perea~\citeyear{Perea12} also provides epistemic
  characterizations of iterated admissibility and EFR that do not
  require complete type structures.}

To summarize, the notion of common knowledge of rationality seems deeply
embedded in many 
game-theoretic
solution concepts.  
While not all solution concepts can be given epistemic
characterizations in terms of some variant of common knowledge of
rationality (one counterexample is the notion of \emph{iterated
regret minimization} \cite{HP11b}), 
the results of this paper and 
of
others mentioned in the previous discussion 
show that
many of the most popular solution concepts do admit such a characterization.

\subsection*{Acknowledgements}
We would like to thank the anonymous reviewers for their detailed
reading of the paper and useful
comments that helped improve the paper. In particular, we thank an
anonymous reviewer for encouraging us to compare our results to those
of Asheim and Perea [2005]. 

\bibliographystyle{chicago}
\bibliography{z,joe}
\end{document}